\documentclass[preprint,superscriptaddress,preprintnumbers,amsmath,amssymb,prd]{revtex4}
\usepackage{graphicx}

\begin{document}

\thispagestyle{empty}

\title{Maximum reflectance and transmittance of films coated with
gapped graphene in the context of Dirac model}

\author{
G.~L.~Klimchitskaya}
\affiliation{Central Astronomical Observatory at Pulkovo of the
Russian Academy of Sciences, Saint Petersburg,
196140, Russia}
\affiliation{Institute of Physics, Nanotechnology and
Telecommunications, Peter the Great Saint Petersburg
Polytechnic University, Saint Petersburg, 195251, Russia}

\author{
V.~M.~Mostepanenko}
\affiliation{Central Astronomical Observatory at Pulkovo of the
Russian Academy of Sciences, Saint Petersburg,
196140, Russia}
\affiliation{Institute of Physics, Nanotechnology and
Telecommunications, Peter the Great Saint Petersburg
Polytechnic University, Saint Petersburg, 195251, Russia}
\affiliation{Kazan Federal University, Kazan, 420008, Russia}

\begin{abstract}
The analytic expressions  for the maximum and minimum reflectance
of optical films coated with gapped graphene
are derived in the application region of the
Dirac model with account of multiple reflections.
The respective film thicknesses are also found. In so doing the
film material is described by the frequency-dependent index of refraction
and graphene by the polarization tensor defined along the real frequency axis.
The developed formalism is illustrated by an example of the graphene-coated
film made of amorphous silica. Numerical computations of the maximum and
minimum reflectances and respective film thicknesses are performed at room
temperature in two frequency regions belonging to the near-infrared and
far-infrared domains. It is shown that in the far-infrared domain the
graphene coating makes a profound effect on the values of maximum reflectance
and respective film thickness leading to a relative increase of their values
by up to 65\% and 50\%, respectively. The maximum transmittance of a
graphene-coated film of appropriately chosen thickness is shown to exceed
90\%. Possible applications of the obtained results are discussed.
\end{abstract}

\maketitle

\section{INTRODUCTION}

The discovery of graphene has already initiated a wide range of both
experimental and theoretical investigations (see, e.g., the monographs
\cite{1,2} and reviews \cite{3,4,5,6}). Graphene is a two-dimensional sheet
of carbon atoms packed in a hexagonal lattice. The most outstanding feature
of graphene is that at energies below 1--2{\ }eV its electronic and
optical properties are well described by the Dirac model. This model assumes
that the quasiparticles in graphene are massless or very light and obey
the linear dispersion relation. Thus, they are described not by the
Schr\"{o}dinger equation, as quasiparticles in all 3D-materials, but by
the Dirac equation, where the speed of light $c$ is replaced with the Fermi
velocity $v_F\approx c/300$ \cite{1,2,3,4,5,6}. In so doing the coupling
constant of quasiparticles to an external electromagnetic field is $e/c$,
where $e$ is the electron charge, as for ordinary electrons.

The electric conductivity, optical properties and closely associated with
them the Casimir effect for graphene sheets have been much investigated using the
Kubo response formalism, the Boltzmann transport equation, the current-current
correlation functions in the random phase approximation, and the two-dimensional
Drude model (see, e.g.,
Refs.~\cite{7,8,9,10,11,12,13,14,15,16,17,18,19,20}).
The most fundamental unified theoretical description of all these phenomena
on the basis of first principles of quantum electrodynamics at nonzero
temperature can be obtained using the polarization tensor of graphene in
(2+1)-dimensional space-time. The explicit expressions for the components of
this tensor at zero temperature were derived in Ref.~\cite{21}.
In Ref.~\cite{22} the polarization tensor for graphene was found at nonzero
temperature, but the values of frequency were restricted to the pure
imaginary Matsubara frequencies. Later another representation for the
polarization tensor, valid over the entire plane of complex frequency, was
obtained in Refs.~\cite{23,24}. The polarization tensor was used
for the theoretical description of the conductivity \cite{25,26,27} and
reflectances \cite{23,28,29,30} of graphene.
The Casimir effect in graphene systems was also investigated in detail using this
approach (see, e.g., Refs.~\cite{31,32,33,34,35,36,37,38,39} and the obtained
theoretical results were found to be in a very good agreement with the measurement
data \cite{40,41}.

It is common knowledge that dielectric films are widely used in optics to
control the amount of light reflected or transmitted at a given wavelength.
In the frequency region, where the refractive index of the film material $n$
is real, the maximum enhancement of
reflected light
at the normal incidence from vacuum
takes place for a film thickness
$d=(2l-1)\lambda/(4n)$, where $\lambda$ is the wavelength of the incident light,
and $l=1,\,2,\,\ldots$ \cite{42}.
Alternatively, the incident light is completely transmitted through a film if
$d=l\lambda/(2n)$ \cite{42}. Taking into consideration that today graphene finds
a growing number of uses, the question arises: In what way
do these classical results have to
 be modified if the optical film is coated with a graphene sheet?
The answer to this question is not trivial because the reflection coefficients
on the surface of graphene are expressed via the polarization tensor and are
quite different from the familiar Fresnel reflection coefficients on a dielectric
surface.

In this paper, we find the  thicknesses of optical films coated with
gapped graphene which ensure either the maximum reflectance or maximum
transmittance of light at a given wavelength with account of multiple
reflections.  This problem is solved
in the framework of the Dirac model using the polarization tensor of graphene.
The analytic expressions for both the maximum reflectance and
transmittance are obtained and compared with the familiar results for uncoated
optical films. Numerical computations are performed for the amorphous
SiO${}_2$ films coated with gapped graphene in the frequency regions where
the imaginary part of SiO${}_2$ index of refraction is equal to zero.
According to our results, graphene coating leads to a relative increase of the
optimal film thickness, providing the maximum value to the reflectance, up to 50\%
in the far-infrared frequency region.
For the SiO${}_2$ film of optimal thickness graphene coating
results in up to 65\% increase of its maximum reflectance.
The film thickness providing the maximum
transmittance of the incident light decreases
under the influence of graphene coating. In so doing,
the maximum transmittance drops from 100\% to about 90\%.

This paper is organized as follows. In Sec.~II we present the general formalism and
derive the analytic expressions for the optimal thicknesses of optical films
and for the maximum reflectances and transmittances in the presence of graphene
coating. Section~III contains the results of numerical computations for
SiO${}_2$ films coated with graphene and comparison with similar results for
uncoated films. In Sec.~IV the reader will find our conclusions and discussion.

\section{Maximum reflectance and transmittance of graphene-coated films}

We consider the dielectric film of thickness $d$ with the index of refraction
$n(\omega)$ in the frequency region where ${\rm Im}n(\omega)=0$.
This dielectric film is coated with a graphene sheet characterized by the gap
of width $\Delta$ in the quasiparticle energy spectrum.
The reflection coefficient for the transverse magnetic (TM), i.e.,
$p$-polarized, electromagnetic wave of frequency $\omega$ on a graphene-coated
plate (regardless of infinite or finite thickness)
takes the form \cite{41}
\begin{equation}
r_{\rm TM}^{(g,f)}(\omega,k)=\frac{r_{\rm TM}^{(g)}(\omega,k)+
r_{\rm TM}^{(f)}(\omega,k)[1-2r_{\rm TM}^{(g)}(\omega,k)]}{1-
r_{\rm TM}^{(g)}(\omega,k)r_{\rm TM}^{(f)}(\omega,k)}.
\label{eq1}
\end{equation}
\noindent
Here, $r_{\rm TM}^{(g)}$ and $r_{\rm TM}^{(f)}$
are the TM reflection coefficients of electromagnetic waves on a graphene
sheet and on a film of thickness $d$ in vacuum, respectively.

In the framework of the Dirac model, the reflection coefficient $r_{\rm TM}^{(g)}$
takes the form \cite{23}
\begin{equation}
r_{\rm TM}^{(g)}(\omega,k)=\frac{p(\omega,k)\Pi_{00}(\omega,k)}{2i\hbar k^2+
p(\omega,k)\Pi_{00}(\omega,k)},
\label{eq2}
\end{equation}
\noindent
where $\Pi_{00}$ is the 00 component of the polarization tensor of graphene,
$p^2=\omega^2/c^2-k^2$, and $k$ is the magnitude of the projection of the
wavevector on the plane of graphene. Note that both the polarization tensor
and the reflection coefficient depend also on the temperature.

It is convenient to introduce the normalized polarization tensor
\begin{equation}
\tilde{\Pi}_{00}(\omega,k)=\frac{p(\omega,k)\Pi_{00}(\omega,k)}{i\hbar k^2}.
\label{eq3}
\end{equation}
\noindent
In terms of $\tilde{\Pi}_{00}$ the reflection coefficient (\ref{eq2})
takes the form
\begin{equation}
r_{\rm TM}^{(g)}(\omega,k)=\frac{\tilde{\Pi}_{00}(\omega,k)}{2+
\tilde{\Pi}_{00}(\omega,k)}.
\label{eq4}
\end{equation}
\noindent
Note that the polarization tensor of graphene is closely connected with the
in-plane conductivity of graphene \cite{25,26,27,35}
\begin{equation}
{\Pi}_{00}(\omega,k)=\frac{4\pi i\hbar k^2}{\omega}
\sigma(\omega,k).
\label{eq5}
\end{equation}
\noindent
Substituting Eq.~(\ref{eq5}) in Eq.~(\ref{eq3}), one finds that
\begin{equation}
\tilde{\Pi}_{00}(\omega,k)=\frac{4\pi p(\omega,k)}{\omega}
\sigma(\omega,k).
\label{eq6}
\end{equation}

{}From Eqs.~(\ref{eq4}) and (\ref{eq6}) we conclude that at the normal incidence
($k=0$)
\begin{equation}
r_{\rm TM}^{(g)}(\omega,0)=\frac{\tilde{\Pi}_{00}(\omega,0)}{2+
\tilde{\Pi}_{00}(\omega,0)},
\label{eq7}
\end{equation}
\noindent
where the quantity
\begin{equation}
\tilde{\Pi}_{00}(\omega,0)=\frac{4\pi}{c}
\sigma(\omega,0)
\label{eq8}
\end{equation}
\noindent
is expressed via the local conductivity of graphene.

The reflection coefficient of the film, $r_{\rm TM}^{(f)}$, at the normal
incidence is well known \cite{42}
\begin{equation}
r_{\rm TM}^{(f)}(\omega,0)=\frac{r(\omega)\left[1-
e^{2i\beta(\omega)}\right]}{1-r^2(\omega)e^{2i\beta(\omega)}},
\label{eq9}
\end{equation}
\noindent
where
\begin{equation}
r(\omega)=\frac{n(\omega)-1}{n(\omega)+1}, \qquad
\beta(\omega)=\frac{2\pi}{\lambda}n(\omega)d.
\label{eq10}
\end{equation}
\noindent
Substituting Eq.~(\ref{eq10}) in Eq.~(\ref{eq9}), we find
\begin{equation}
r_{\rm TM}^{(f)}(\omega,0)=
\frac{[n^2(\omega)-1]\sin\beta}{[n^2(\omega)+1]\sin\beta+
2in(\omega)\cos\beta}.
\label{eq11}
\end{equation}

Now we are in a position to obtain explicit expressions for the reflection
coefficients and reflectance on the graphene-coated film at the normal
incidence
\begin{eqnarray}
&&
r^{(g,f)}(\omega)=|r_{\rm TM}^{(g,f)}(\omega,0)|=
|r_{\rm TE}^{(g,f)}(\omega,0)|,
\nonumber\\
&&
{\cal R}^{(g,f)}(\omega)={r^{(g,f)}}^2(\omega),
\label{eq12}
\end{eqnarray}
\noindent
where the transverse electric, i.e., $s$ polarization is notated as TE.
For this purpose we introduce the notation
\begin{equation}
\tilde{\Pi}_{00}(\omega,0)=a(\omega)+ib(\omega)
\label{eq13}
\end{equation}
\noindent
and calculate the quantities (\ref{eq12}) by substituting Eqs.~(\ref{eq7})
and (\ref{eq11}) in Eq.~(\ref{eq1}) taken at $k=0$
\begin{widetext}
\begin{eqnarray}
&&
r^{(g,f)}=\left|\frac{(n^2+a-1)\sin\beta-bn\cos\beta+
i(b\sin\beta+an\cos\beta)}{(n^2+a+1)\sin\beta-bn\cos\beta+
i[b\sin\beta+n(a+2)\cos\beta]}
\right|,
\nonumber\\
&&
{\cal R}^{(g,f)}=\frac{[(n^2+a-1)\sin\beta-bn\cos\beta]^2+
(b\sin\beta+an\cos\beta)^2}{[(n^2+a+1)\sin\beta-bn\cos\beta]^2+
[b\sin\beta+n(a+2)\cos\beta]^2}.
\label{eq14}
\end{eqnarray}
\end{widetext}
\noindent
Here and sometimes below, for the sake of brevity, we omit the argument $\omega$
in the quantities $r^{(g,f)}$, ${\cal R}^{(g,f)}$, $n$, $a$, $b$, and $\beta$.

The explicit expressions for the real and imaginary parts of $\tilde{\Pi}_{00}$,
due to Eq.~(\ref{eq8}), differ from the real and imaginary parts of the local
conductivity of graphene by only the factor $4\pi/c$. Because of this, using
the results for conductivity from Ref.~\cite{26}, one obtains
\begin{eqnarray}
&&
a=\alpha\pi\theta(\hbar\omega-\Delta)
\frac{(\hbar\omega)^2+\Delta^2}{(\hbar\omega)^2}\tanh\frac{\hbar\omega}{4k_BT},
\label{eq15} \\
&&
b=\alpha\left[\frac{2\Delta}{\hbar\omega}-
\frac{(\hbar\omega)^2+\Delta^2}{(\hbar\omega)^2}
\ln\left|\frac{\hbar\omega+\Delta}{\hbar\omega-\Delta}\right|
+Y(\omega,T)\right],
\nonumber
\end{eqnarray}
\noindent
where
\begin{equation}
Y(\omega,T)=\frac{8c}{\omega}\int_{\frac{\Delta}{2\hbar c}}^{\infty}
\frac{du}{\exp\frac{\hbar cu}{k_BT}+1}\,
\frac{4(\hbar cu)^2+\Delta^2}{4(\hbar cu)^2-(\hbar\omega)^2}.
\label{eq16}
\end{equation}
\noindent
Here, $\alpha=e^2/(\hbar c)$ is the fine-structure constant,
$\Delta$ is the width of the gap in the spectrum of quasiparticles
(usually $\Delta$ is very small but for graphene on a substrate may
reach 0.2\,eV \cite{3,13,21}), and $\theta(x)$ is the step function
equal to unity for $x\geq 0$ and to zero otherwise.

It is common knowledge \cite{42} that the reflectance of an uncoated optical
film under the normal incidence,
\begin{equation}
{\cal R}^{(f)}(\omega)=|r_{\rm TM}^{(f)}(\omega,0)|^2=
|r_{\rm TE}^{(f)}(\omega,0)|^2,
\label{eq17}
\end{equation}
\noindent
where $r_{\rm TM}^{(f)}(\omega,0)$ is defined in Eq.~(\ref{eq11}), reaches
the maximum value under the condition $|\sin\beta|=1$, $\cos\beta=0$, i.e.,
for the film thickness
\begin{equation}
d_{\max}=\frac{\lambda}{4n}(2l-1),\quad
l=1,\,2,\,\ldots\, .
\label{eq18}
\end{equation}
\noindent
The maximum value of the reflectance is equal to \cite{42}
\begin{equation}
{\cal R}_{\max}^{(f)}(\omega)=\left[
\frac{n^2(\omega)-1}{n^2(\omega)+1}\right]^2.
\label{eq19}
\end{equation}

The minimum value of the reflectance is equal to zero. It is reached under
the condition $\sin\beta=0$, i.e., for the film thickness
\begin{equation}
d_{\min}=\frac{\lambda}{2n}l.
\label{eq19a}
\end{equation}
\noindent
In this case the wave is completely transmitted through the film.

Now we derive the values of film thickness and of the maximum reflectance and
transmittance for a film coated with graphene. For this purpose, it is
convenient to rewrite Eq.~({\ref{eq14}) for the reflectance ${\cal R}^{(g,f)}$
in an equivalent form
\begin{equation}
{\cal R}^{(g,f)}=
\frac{A\cos^2\beta+B\sin^2\beta+C\cos\beta\sin\beta}{D\cos^2\beta
+E\sin^2\beta+C\cos\beta\sin\beta},
\label{eq20}
\end{equation}
\noindent
where
\begin{eqnarray}
&&
A=n^2(a^2+b^2),\quad B=(a+n^2)^2+b^2+1-2a-2n^2,
\nonumber \\
&&
C=2bn(1-n^2),\quad D=n^2[(a+2)^2+b^2],
\nonumber \\
&&
E=(a+n^2)^2+b^2+1+2a+2n^2.
\label{eq21}
\end{eqnarray}

We, next, consider the condition
\begin{equation}
\frac{\partial{\cal R}^{(g,f)}}{\partial\beta}=0,
\label{eq22}
\end{equation}
\noindent
which ensures the extremum value of ${\cal R}^{(g,f)}$.
The condition (\ref{eq22}) results in the equation
\begin{equation}
\tan^2\beta+\frac{2p}{r}\tan\beta+\frac{q}{r}=0.
\label{eq23}
\end{equation}
\noindent
Here, the following notations are introduced:
\begin{eqnarray}
&&
p=BD-AE=4n^2(n^2-1)(a^2-b^2+an^2+n^2+a-1),
\nonumber\\
&&
q=C(D-A)=-8bn^3(n^2-1)(a+1),
\nonumber\\
&&
r=C(B-E)=8bn(n^2-1)(a+n^2).
\label{eq24}
\end{eqnarray}

{}From Eq.~(\ref{eq23}) one obtains
\begin{equation}
\tan\beta_{\pm}=-\frac{p}{r}\pm\sqrt{\frac{p^2}{r^2}-\frac{q}{r}}.
\label{eq25}
\end{equation}
\noindent
Note that according to Eq.~(\ref{eq24}) the quantity $q/r$ is always negative
independently on the sign of $b$.

Equations (\ref{eq10}) and (\ref{eq25}) allow finding the values of film thickness
which ensure the maximum and minimum values of the reflectance. The resolution of this
issue depends on the sign of the imaginary part $b$ of the polarization tensor
$\tilde{\Pi}_{00}$ which, in its turn, depends on the values of frequency and
temperature (see Sec.~III).
If $b>0$, than from Eq.~(\ref{eq24}) it follows that $r>0$ because the real part
$a$ of  $\tilde{\Pi}_{00}$ is nonnegative due to Eq.~(\ref{eq15}).
Let us now assume that $p>0$ as well. In this case $p/r>0$ and from Eq.~(\ref{eq25})
we have $\tan\beta_{-}<0$ and $|\tan\beta_{-}|>\tan\beta_{+}$.
Then from Eqs.~(\ref{eq10}) and (\ref{eq25}) one obtains
\begin{equation}
d_{-}=\frac{\lambda}{2\pi n}\left[\pi+
\arctan\left(-\frac{p}{r}-\sqrt{\frac{p^2}{r^2}-\frac{q}{r}}\right)\right].
\label{eq26}
\end{equation}
\noindent
It is easily seen that in this case $d_{-}$ is the smallest film
thickness which ensures the maximum value of the
reflectance. At the same time, for the graphene-coated film of
the smallest thickness
\begin{equation}
d_{+}=\frac{\lambda}{2\pi n}
\arctan\left(-\frac{p}{r}+\sqrt{\frac{p^2}{r^2}-\frac{q}{r}}\right),
\label{eq27}
\end{equation}
\noindent
the reflectance takes the minimum value.
It may be also that the quantity $p$ defined in
Eq.~(\ref{eq24}) is opposite in sign, i.e., $p<0$. In this situation the meaning of
$d_{-}$ and $d_{+}$ interchange, i.e., $d_{-}$ from Eq.~(\ref{eq26}) ensures
the minimum value of the reflectance and $d_{+}$ from Eq.~(\ref{eq27}) corresponds
to the maximum reflectance.

Now we consider the case $b<0$ and $p>0$. In this case it follows
from Eq.~(\ref{eq24}) that $r<0$ and $p/r<0$.
In such a manner one obtains  from Eq.~(\ref{eq25}) that
$\tan\beta_{+}>0$, $\tan\beta_{-}<0$ and
$\tan\beta_{+}>|\tan\beta_{-}|$.
Then $d_{+}$ is again given by Eq.~(\ref{eq27}) but corresponds to the maximum
value of the reflectance, and $d_{-}$ is given by
Eq.~(\ref{eq26}) and corresponds to the minimum reflectance.
If the condition $p<0$ holds, the meaning of $d_{-}$ and $d_{+}$ interchange
again.

The extremum values of the reflectance of a graphene-coated film are now
determined from Eq.~(\ref{eq20})
\begin{equation}
{\cal R}_{\pm}^{(g,f)}=
\frac{A+B\tan^2\beta_{\pm}+C\tan\beta_{\pm}}{D
+E\tan^2\beta_{\pm}+C\tan\beta_{\pm}},
\label{eq28}
\end{equation}
\noindent
where $\tan\beta_{\pm}$ is presented in Eq.~(\ref{eq25}).
In so doing the extremum values of the transmittance are given by
\begin{equation}
T_{\pm}^{g,f)}=1-{\cal R}_{\pm}^{(g,f)}.
\label{eq29}
\end{equation}
\noindent
In the next section the above results are illustrated by a specific example.

\section{Graphene-coated silica films}

Here, we perform numerical computations of the optimal thicknesses and maximum
reflectances and transmittances of amorphous SiO${}_2$ films coated with
gapped graphene at room temperature ($T=300{\ }$K).
In the application region of Dirac model, there are two frequency regions where
${\rm Im}n=0$. The first of them belongs to the near-infrared domain. In this region
$\hbar\omega$ varies from 0.35 to 1{\ }eV and the respective wavelength changes
from 3.54 to $1.24{\ }\mu$m. Below we demonstrate that for these frequencies of
incident light the graphene coating makes only a minor impact on the optical
properties of a film.

Thus, for an uncoated film the optimal thicknesses (\ref{eq18}) and (\ref{eq19a}),
ensuring the maximum and minimum reflectance for $l=1$, vary from 0.630 to
$0.214{\ }\mu$m and from 1.260 to
$0.428{\ }\mu$m, respectively.
The respective maximum reflectance ${\cal R}_{\max}^{(f)}$ calculated using
Eq.~(\ref{eq19}) varies from 0.1075 to 0.1258. The maximum transmittance  is
equal to unity. Here and below we use the values of $n(\omega)$ for SiO${}_2$
from Ref.~\cite{43}.

Now we consider the graphene-coated film within the same frequency region.
The real part of $\tilde{\Pi}_{00}$, $a$, is given by the first expression in
Eq.~(\ref{eq15}) where $\Delta<\hbar\omega$ for all realistic values of
$\Delta$ (see Sec.~II). The imaginary part of $\tilde{\Pi}_{00}$, $b$, is
given by the second line in Eq.~(\ref{eq15}).
Note that at room temperature $k_BT=0.026{\ }$eV, i.e., $k_BT\ll\hbar\omega$
in the frequency region under consideration. If, in addition, the condition
$\Delta\ll k_BT$ is satisfied, the quantity $b$ can be presented in a simpler
form \cite{26}
\begin{equation}
b=\alpha\left[\frac{2\Delta}{\hbar\omega}-
\frac{(\hbar\omega)^2+\Delta^2}{(\hbar\omega)^2}
\ln\left|\frac{\hbar\omega+\Delta}{\hbar\omega-\Delta}\right|
-48\zeta(3)\left(\frac{k_BT}{\hbar\omega}\right)^3\right],
\label{eq30}
\end{equation}
\noindent
where $\zeta(z)$ is the Riemann zeta function. For the gapless graphene
$\Delta=0$ and Eq.~(\ref{eq30}) is simplified to
\begin{equation}
b=-48\alpha\zeta(3)\left(\frac{k_BT}{\hbar\omega}\right)^3.
\label{eq31}
\end{equation}
\noindent
Under the opposite condition $\Delta\gg k_BT$ Eq.~(\ref{eq15}) for $b$
results in \cite{26}
\begin{equation}
b=\alpha\left[\frac{2\Delta}{\hbar\omega}-
\frac{(\hbar\omega)^2+\Delta^2}{(\hbar\omega)^2}
\ln\left|\frac{\hbar\omega+\Delta}{\hbar\omega-\Delta}\right|
-24\frac{k_BT\Delta^2}{(\hbar\omega)^3}
e^{-\frac{\Delta}{2k_BT}}\right].
\label{eq32}
\end{equation}

{}From Eqs.~(\ref{eq30})--(\ref{eq32}) it is easily seen that $b<0$ and its
magnitude is negligibly small. Thus, for $\Delta=0$
the quantity $b$ varies from
$-1.7\times 10^{-4}$ to $-7.3\times 10^{-6}$ when $\hbar\omega$ varies from
0.35 to 1{\ }eV, whereas $a\approx 0.0229$ and does not depend on $\omega$.
For $\Delta\neq 0$ the magnitudes of $b$ are even less. Then, from Eq.~(\ref{eq24})
we find that $p>0$ holds and, according to Sec.~II, $\beta_{+}$ and
$\beta_{-}$ correspond to the maximum and minimum reflectances, respectively.
Performing calculations by Eqs.~(\ref{eq26}) and (\ref{eq27}) in the case of
$\Delta=0$, one obtains that in the frequency region under consideration $d_{+}$
decreases from 0.6299 to $0.2139{\ }\mu$m and $d_{-}$
decreases from 1.2599 to $0.4279{\ }\mu$m.
These coincide with the above results for an uncoated film.
The values of the maximum and minimum reflectances are calculated using Eq.~(\ref{eq20}).
Thus, in the considered frequency region ${\cal R}_{+}^{(g,f)}$ varies from 0.1109
to 0.1292 and ${\cal R}_{-}^{(g,f)}$ is equal to $1.28\times 10^{-4}$ and does
not depend on frequency. It is seen that the presence of graphene coating leads to
only minor increase of ${\cal R}_{+}^{(g,f)}$ and has almost no impact on
${\cal R}_{-}^{(g,f)}$. With increasing $\Delta$ the above results remain
unchanged.

We come now to the second frequency region where the imaginary part of
$n$ for SiO${}_2$ is negligibly small. This is the region from 0.62 to
12.4{\ }meV belonging to the far-infrared domain. The corresponding wavelength
varies from 100 to $2000{\ }\mu$m and $n$ increases from 1.955 to 1.967 \cite{43}.
Here, the impact of graphene coating on the optimal values of film thickness
and maximum and minimum reflectances increases with decreasing frequency and becomes
rather large at $\hbar\omega<5{\ }$meV.

Computations are performed with the exact equations (\ref{eq15}), (\ref{eq16}),
(\ref{eq21}), (\ref{eq24}) and (\ref{eq26})--(\ref{eq28}). We first consider
the values of the gap width larger than all considered frequencies, i.e.,
$\Delta>\hbar\omega$ (remind that at room temperature $\hbar\omega<k_BT$).
From Eq.~(\ref{eq15}) it is easily seen that in the frequency region under
consideration $a=0$, $b>0$, and $p>0$.  Thus, according to the results of Sec.~II,
$d_{-}$ in Eq.~(\ref{eq26}) ensures the maximum value of the reflectance and
$d_{+}$ in Eq.~(\ref{eq27}) leads to the minimum reflectance (i.e., to the
maximum transmittance).

The computational results for the reflectances are presented in Fig.~\ref{fg1}
as functions of frequency of the incident light. The pair of solid lines notated 1
shows reflectances of graphene-coated silica films ${\cal R}_{-}^{(g,f)}$ which
thicknesses $d_{-}$ are determined from the condition of maximum reflectance.
The pair of solid lines notated 2
shows reflectances of a graphene-coated film ${\cal R}^{(g,f)}$ which
thickness $d_{\max}$ is found from the condition of maximum reflectance
(\ref{eq18}) with no account of graphene.
In each pair the top line is plotted for graphene with $\Delta=15{\ }$meV and
the bottom line  for graphene with $\Delta=50{\ }$meV.
The dashed line demonstrates the maximum reflectance of an uncoated film.
It is almost flat because in the frequency region considered $n$ has only
minor dependence on $\omega$.

As seen in Fig.~\ref{fg1}, in the far-infrared region of the spectrum the
presence of graphene coating significantly affects the maximum reflectance of
a silica film. The largest impact is reached at $\hbar\omega=0.62{\ }$meV.
Here, the optimal film thickness is equal to $d_{-}=378.98{\ }\mu$m for
$\Delta=15{\ }$meV and  ${\cal R}_{-}^{(g,f)}=0.5494$ (the top line of pair 1).
This should be compared with the optimal thickness $d_{\max}=255.72{\ }\mu$m
of an uncoated film leading to its maximum reflectance.
The film of thickness $d_{\max}$, if coated with graphene, would have a
reflectance of only ${\cal R}^{(g,f)}=0.4057$. The maximum reflectance of an
uncoated film is equal to ${\cal R}_{\max}^{(f)}=0.3425$.
Thus, a relative increase in the maximum reflectance due to graphene coating
in this frequency range is equal to  60\%. This is reached by a relative
increase of the optimal film thickness by 48\%.
The minimum reflectance of a graphene-coated film at $\hbar\omega=0.62{\ }$meV
is equal to ${\cal R}_{+}^{(g,f)}=0.07588$. It is reached for a film of
$d_{+}=123.26{\ }\mu$m thickness. This should be compared with
${\cal R}_{\min}^{(f)}=0$ and $d_{\min}=511.44{\ }\mu$m.
Note, however, that the transmittance of the film of thickness $d_{+}$ reaches
0.924, i.e., remains rather high.

Next we consider the case when the width of the gap is smaller than all
considered frequencies, i.e., $\Delta<\hbar\omega$
(remind that  $\hbar\omega<k_BT$ at $T=300{\ }$K).
Although in this case $a\neq 0$, the inequalities $b>0$ and $p>0$ remain
valid. Thus, $d_{-}$ in Eq.~(\ref{eq26})  and
$d_{+}$ in Eq.~(\ref{eq27}) again ensure that the reflectances
${\cal R}_{-}^{(g,f)}$ and ${\cal R}_{+}^{(g,f)}$ reach the maximum and
minimum values, respectively.

In Fig.~\ref{fg2} we present the computational results for the reflectances
as functions of frequency. Similar to Fig.~\ref{fg1},
the pair 1 of solid lines
shows reflectances  ${\cal R}_{-}^{(g,f)}$ for the films of
thickness $d_{-}$ and the pair  2
shows reflectances  ${\cal R}^{(g,f)}$ for the films of
thickness $d_{\max}$ found with no account of graphene.
The top and bottom lines in each pair are for graphene coating
with $\Delta=0$ and 0.6{\ }meV, respectively. The region of small
frequencies is shown on an inset for better visualization.
The dashed line is for an uncoated film.

As can be seen in Fig.~\ref{fg2}, graphene coating with a smaller
width of the band gap strongly affects the maximum reflectances although
the impact of a specific value
of $\Delta$ is not so pronounced as in Fig.~\ref{fg1}.
Thus, for a gapless graphene coating at $\hbar\omega=0.62{\ }$meV we have
$d_{-}=383.9{\ }\mu$m and ${\cal R}_{-}^{(g,f)}=0.5686$ (the top line of pair 1).
At the same time, if a graphene-coated film has the thickness
$d_{\max}=255.7{\ }\mu$m determined with no account of graphene we obtain the
smaller value ${\cal R}^{(g,f)}=0.4142$ (the top line of pair 2).
The maximum reflectance of an uncoated film in this case
${\cal R}_{\max}^{(f)}=0.3425$ is reached for a film of thickness
$d_{\max}=255.7{\ }\mu$m.
The resulting relative increase in the maximum reflectance due to the
presence of graphene coating
is equal to  66\%. The respective relative change in
the film thickness is by 50.1\%. At $\hbar\omega=0.62{\ }$meV
the minimum reflectance of a graphene-coated film
${\cal R}_{+}^{(g,f)}=0.091$ is reached for a film  thickness
$d_{+}=128.2{\ }\mu$m. For an uncoated film
${\cal R}_{\min}^{(f)}=0$ is reached for $d_{\min}=511.44{\ }\mu$m.
In this case the maximum transmittance is equal to
0.91.

The last possibility, when $\Delta$ is larger than
$\min(\hbar\omega)$ in our frequency region but less than
$\max(\hbar\omega)$, remains to be investigated. Here, again, $b>0$
with exception of only a vicinity of frequency $\hbar\omega=\Delta$,
where $b$ goes to $-\infty$. In the frequency region under consideration
the width of this vicinity is unphysically narrow \cite{26}.
As a result,  $d_{-}$ and
$d_{+}$ in Eqs.~(\ref{eq26})  and (\ref{eq27}) provide the maximum and
minimum values to the reflectance of a graphene-coated
film, respectively.

The computational results for the reflectances as functions of frequency are
presented in Fig.~\ref{fg3}. The top solid line shows the maximum reflectance
of a SiO${}_2$ film coated by graphene with $\Delta=3{\ }$meV, where the
optimal film thickness $d_{-}$  is determined with account of the graphene
coating. The bottom solid line shows similar results but for a film thickness
$d_{\max}$ found with no account of graphene. As in Figs.~\ref{fg1} and \ref{fg2},
the dashed line reproduces the maximum reflectance of an uncoated film of
thickness $d_{\max}$. As an example, for $\hbar\omega=0.62{\ }$meV we have
${\cal R}_{-}^{(g,f)}=0.5673$ (the top line) and $d_{-}=383.5{\ }\mu$m.
At the same frequency the maximum reflectance for a film of thickness
$d_{\max}=255.7{\ }\mu$m is equal to ${\cal R}^{(g,f)}=0.4136$ (the bottom line).
In the absence of the graphene coating we again have ${\cal R}_{\max}^{(f)}=0.3425$.
This leads to the maximum relative change of 65.6\% due to the
presence of graphene. The minimum reflectance
${\cal R}_{+}^{(g,f)}=0.091$ is reached for a film  thickness
$d_{+}=127.8{\ }\mu$m. This corresponds to the transmittance of 91\%.
Note also the presence of the vertical peak at $\hbar\omega=\Delta$.
It is caused by the break of continuity where $b$ goes to $-\infty$ and,
thus, ${\cal R}=1$. As mentioned above, this peak in the reflectance is
unphysically narrow and only part of it is shown in Fig.~\ref{fg3}.

\section{Conclusions and discussion}

In the foregoing, we have investigated the impact of graphene coating on the
maximum and minimum reflectances of transparent optical films at normal
incidence with account of multiple reflections.
The general analytic expressions are found for the film thicknesses
leading to the maximum and minimum reflectances in the presence of graphene.
These results are determined by the index of refraction of film material
(as for uncoated films in the commonly known classical case) and by the 00 component
of the polarization tensor of graphene found recently in the framework of the
Dirac model. Thus, the respective film thicknesses and the maximum and
minimum reflectances and transmittances can now be calculated for any film
material at any temperature using the developed formalism.

The obtained results were illustrated with the example of amorphous SiO${}_2$
films which are often used as the underlying substrates for graphene sheets
(see, e.g., Refs.~\cite{40,44}). It has been known that in the application
region of the Dirac model this material possesses two frequency regions of
transparency belonging to the near-infrared and far-infrared domains of the
spectrum. We have calculated both the maximum and minimum reflectances and
respective film thicknesses for films coated with graphene with different
values of the gap width and compared them with respective values for an
uncoated film. According to our results, in the near-infrared domain graphene
makes only a minor impact on the reflectance and transmittance of an uncoated
film.

Quite to the opposite, it was shown that for frequencies belonging to the
far-infrared domain graphene coating makes a profound effect on the
maximum and minimum reflectances, and also on the thicknesses of films
whereby these extremum values are reached. Here, we have considered three
options depending on the relationship between the width of a graphene gap
and the frequency. In all cases considered the relative increase of the maximum
reflectance due to graphene coating is up to 65\%. This is accompanied with
up to 50\% relative increase of the film thickness whereby the maximum
reflectance is reached. It was shown also that the graphene-coated SiO${}_2$
film of appropriate thickness may possess a transmittance of more than 90\%,
i.e., for less than 10\% smaller than that in the absence of graphene coating.

The obtained results are interesting not only for fundamental physics.
At the present time graphene-coated films made of different materials find
prospective technological applications in optoelectronic devices \cite{45,46},
corrosion protection \cite{47,48}, solar cells \cite{49,50}, transparent
electrodes \cite{51}, biosensors \cite{52}, and in many others.
These applications would require the greatest possible transmittance
or reflectance of the graphene-coated films in use.
The optical parameters of such films can be specified beforehand using
the formalism developed.

%%%%%%%%%%%%%%%%%%%%%%%%%%%%%%%%%%%%%%%%%%%%%%%%%%%%%%%%%%%%%%%%%%%%%%%%%%%%%%%%
\section*{Acknowledgments}
The work of V.M.M.\ was partially supported by the Russian
Government
Program of Competitive Growth of Kazan Federal University.
%%%%%%%%%%%%%%%%%%%%%%%%%%%%%%%%%%%%%%%%%%%%%%%%%%%%%%%%%%%%%%%%%%%%%%%%%%%%%%%%

%\end{document}
%%%%%%%%%%%%%%%%%%%%%%%%%%%%%%%%%%%%%%%%%%%%%%%%%%%%%%%%%%%%%%%%%%%%%%
\newpage
%%%%%%%__FIGURE__1__%%%%%%%%%%%%%%%%%%%%
\begin{figure}[b]
\vspace*{-11cm}
\centerline{\hspace*{2.5cm}
\includegraphics{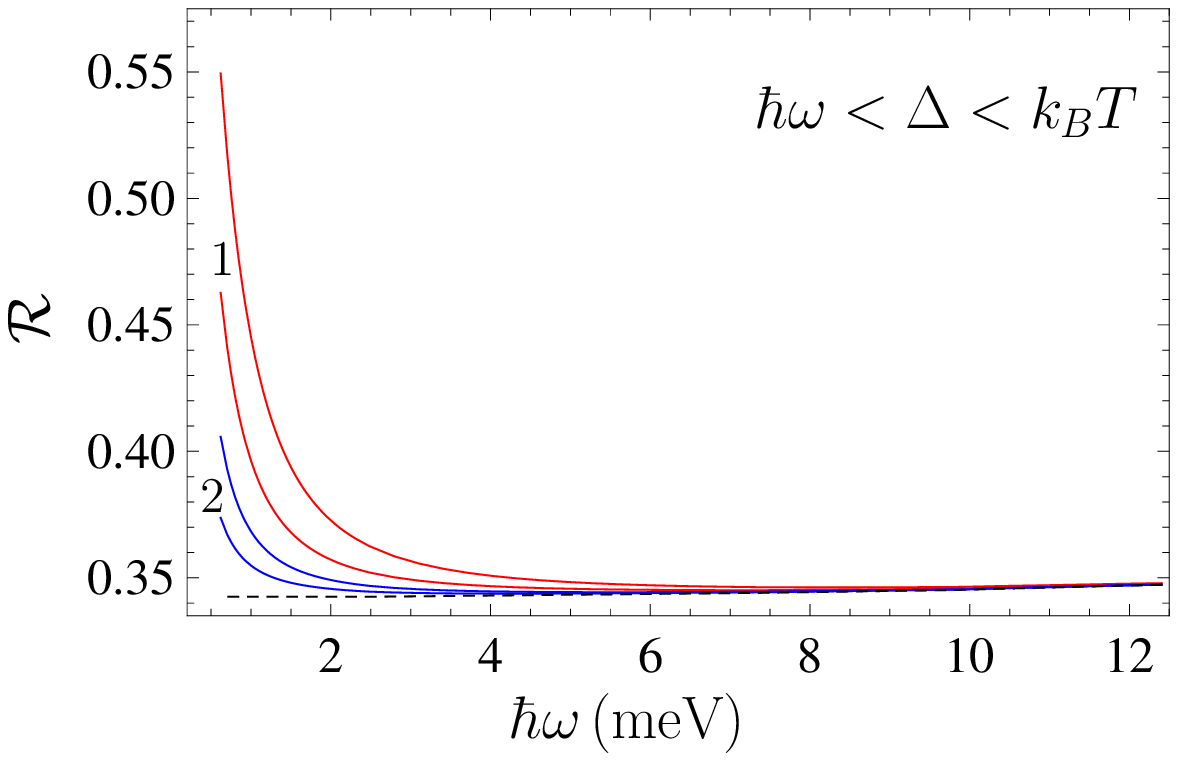}
}
\vspace*{-9.5cm}
\caption{\label{fg1}
The pairs of solid lines 1 and 2 show the reflectances of graphene-coated
SiO${}_2$ films as functions of frequency under the condition that the film
thickness is found from the demand of maximum reflectance with and
without a graphene coating, respectively. In each pair of lines the
top one is for $\Delta=15{\ }$meV and the bottom one is for $\Delta=50{\ }$meV.
The dashed line shows the maximum reflectance of an uncoated film.
}
\end{figure}
%%%%%%%%%%%%%
%%%%%%%__FIGURE__2__%%%%%%%%%%%%%%%%%%%%
\begin{figure}[b]
\vspace*{-11cm}
\centerline{\hspace*{2.5cm}
\includegraphics{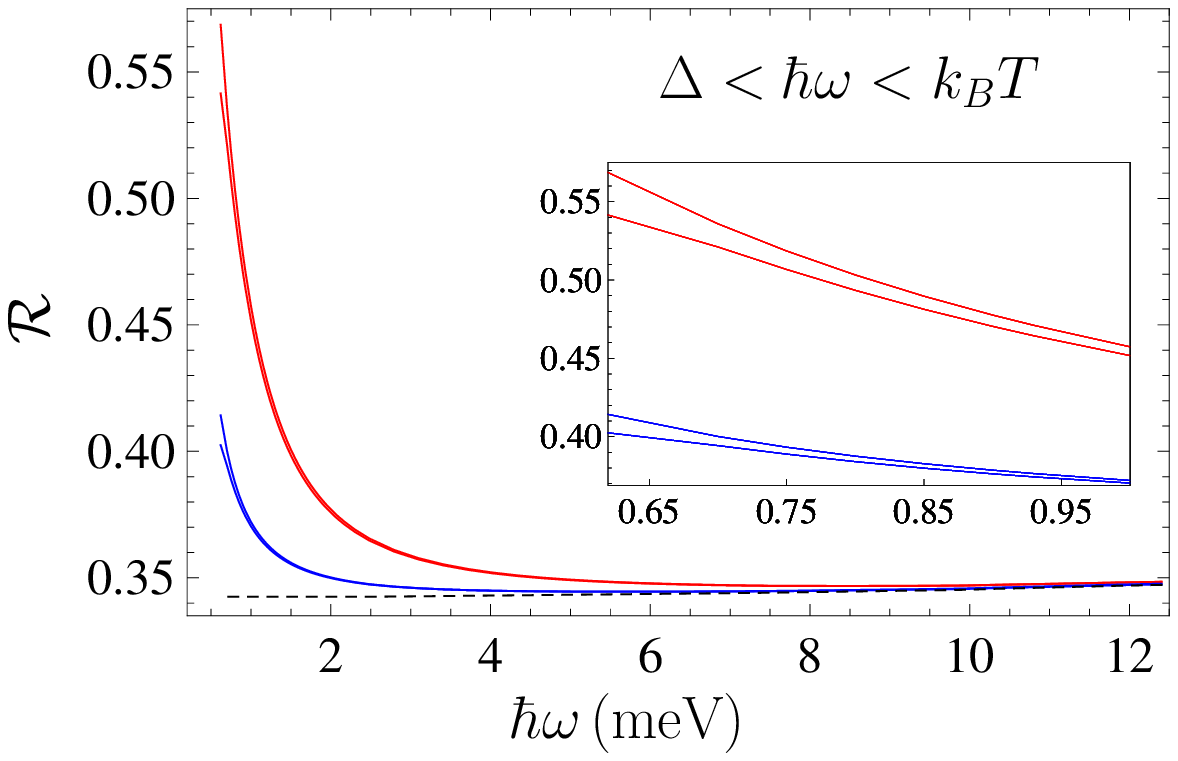}
}
\vspace*{-9.5cm}
\caption{\label{fg2}
The pairs of solid lines 1 and 2 show the reflectances of graphene-coated
SiO${}_2$ films as functions of frequency under the condition that the film
thickness is found from the demand of maximum reflectance with and
without  a graphene coating, respectively. In each pair of lines the
top one is for $\Delta=0$ and the bottom one is for $\Delta=0.6{\ }$meV.
The dashed line shows the maximum reflectance of an uncoated film.
The region of small frequencies is presented in an inset on a larger scale.
}
\end{figure}
%%%%%%%%%%%%%
%%%%%%%__FIGURE__3__%%%%%%%%%%%%%%%%%%%%
\begin{figure}[b]
\vspace*{-11cm}
\centerline{\hspace*{2.5cm}
\includegraphics{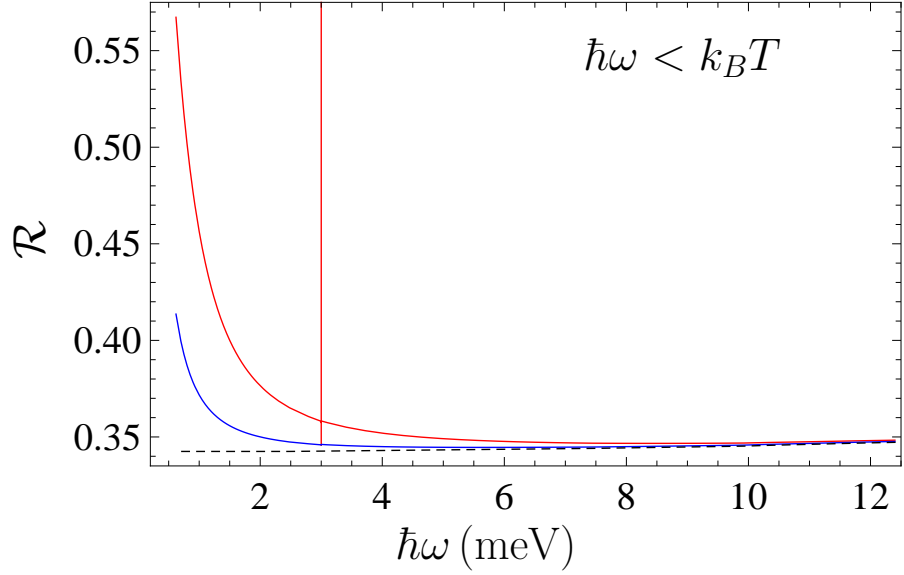}
}
\vspace*{-9.5cm}
\caption{\label{fg3}
The top and bottom solid lines show the reflectances of SiO${}_2$ films
coated by graphene with $\Delta=3{\ }$meV
as functions of frequency under the condition that the film
thickness is found from the demand of maximum reflectance with and
without a graphene coating, respectively.
The dashed line shows the maximum reflectance of an uncoated film.
The vertical line (which should be extended to unity) corresponds to the break of
continuity in the imaginary part of the polarization tensor at
$\hbar\omega=\Delta$.
}
\end{figure}
%%%%%%%%%%%%%
%%%%%%%%%%%%
\end{document}